\documentclass{emulateapj}

\usepackage{epstopdf}

\newcommand{\Msun}{$M_{\sun}$}

\shorttitle{R-PROCESS NUCLEOSYNTHESIS IN NEUTRON STAR MERGERS}
\shortauthors{Goriely et al.}

\begin{document}

\title{R-PROCESS NUCLEOSYNTHESIS IN DYNAMICALLY EJECTED MATTER OF NEUTRON STAR MERGERS}

\author{Stephane Goriely}
\affil{Institut d'Astronomie et d'Astrophysique, Universit\'e Libre de Bruxelles, C.P. 226, B-1050 Brussels,
Belgium}
%{sgoriely@astro.ulb.ac.be}
\author{Andreas Bauswein and Hans-Thomas Janka}
\affil{Max-Planck-Institut f\"ur Astrophysik, Postfach 1317, D-85741 Garching, Germany}

\begin{abstract}

Although the rapid neutron-capture process, or r-process, is 
fundamentally important for explaining the origin of approximately half
of the stable nuclei with $A > 60$, the astrophysical site of
this process has not been identified yet. Here we study r-process nucleosynthesis
in material that is dynamically ejected by tidal and pressure forces during
the merging of binary neutron stars (NSs) and within milliseconds afterwards.
For the first time we make use of relativistic hydrodynamical simulations of such
events, defining consistently the conditions that determine the nucleosynthesis,
i.e., neutron enrichment, entropy, early density evolution and thus expansion 
timescale, and ejecta mass. We find that $10^{-3}$--$10^{-2}\,M_\odot$ are ejected,
which is enough for mergers to be the main source of heavy ($A \gtrsim 140$) galactic
r-nuclei for merger rates of some $10^{-5}$yr$^{-1}$.
While asymmetric mergers eject 2--3 times more mass than symmetric ones, the 
exact amount depends weakly on whether the NSs have radii of $\sim$15\,km for a
``stiff'' nuclear equation of state (EOS) or $\sim$12\,km for a ``soft'' EOS.
R-process nucleosynthesis during the decompression becomes
largely insensitive to the detailed conditions because of efficient fission
recycling, producing a composition that closely
follows the solar r-abundance distribution for nuclei
with mass numbers $A > 140$. Estimating the light curve powered by the radioactive
decay heating of r-process nuclei with an approximative model, we expect high
emission in the B-V-R bands for 1--2 days with potentially observable longer 
duration in the case of asymmetric mergers because of the larger ejecta mass.

\end{abstract}

\today

\section{INTRODUCTION}

The r-process, or rapid neutron-capture process, of stellar nucleosynthesis is
invoked to explain the production of the stable (and some long-lived radioactive) 
neutron-rich nuclides heavier than iron, which are observed in stars of different
metallicities as well as in the solar system \citep[for a review, see][]{ar07}.
Despite a growing wealth of observational data \citep[e.g.,][]{sneden08} 
and although increasingly better r-process
models with new astrophysical or nuclear physics ingredients
have been developed over decades, the stellar production site(s) of r-process
material has (have) not been identified yet. All proposed scenarios face serious
problems. Supernovae (SNe), for example, appear attractive
because of their potential to explain observational features of the galactic
chemical enrichment history \citep[e.g.,][]{argast04}.
Their nucleosynthesis, however, exhibits extreme sensitivity to the detailed
conditions in the ejecta, whose viability for strong r-processing could not be
verified by sophisticated hydrodynamical models
\citep[e.g.,][]{hoffman08,janka08,roberts10,hudepohl10,fisch10,wanajo11}.

Early in the development of the theory of nucleosynthesis, an alternative origin of
r-process nuclei was proposed \citep{tsuruta65}.
It relies on the fact that at high densities (typically $\rho > 10^{10}$ gcm$^{-3}$) 
matter tends to be composed of nuclei lying on the neutron-rich side of the valley 
of nuclear stability as a result of endothermic free-electron captures.
The astrophysical plausibility of this production mechanism as the source of the
observed r-nuclides has long been questioned. It remained largely unexplored until 
\citet{lattimer77} and \citet{meyer89}
studied the decompression of cold, neutronized matter ejected by tidal effects 
of a black hole (BH) on a neutron star (NS) companion.
Recently, special attention has been paid to NS mergers because 
hydrodynamic simulations of NS-NS and NS-BH mergers showed that a non-negligible 
amount of matter may be ejected \citep[e.g.,][]{janka99,ros04,oech07}.
%  The decompression and ejection of initially cold NS matter might also happen in 
%  other astrophysical scenarios like giant flares in soft-gamma repeaters \citep{granot06},
%  the explosion of a NS eroded below its minimum mass \citep{sumi98},
%  or the equatorial shedding of material from very rapidly rotating supermassive or 
%  hypermassive NSs \citep[for more details, see][]{ar07}.

All previous investigations of the ejecta from coalescing NSs
\citep{frei99,go05,ar07,metzger10,roberts11} were parametrized in one way or 
another, which makes their results and conclusions subject to open questions. 
In \citet{go05} and \citet{ar07} the thermodynamic profiles were constructed by a 
simple decompression model \citep[see][]{go11}, 
but the neutron enrichment (or equivalently the electron fraction $Y_e$) was 
consistently taken from $\beta$-equilibrium assumed to have been achieved at the
initial density prior to the decompression. It was found that the final 
composition of the material ejected from the inner crust depends on the initial 
density, at least for the outer parts of the inner crust at 
$\rho_{\mathrm{drip}} \le \rho \le 10^{12}$g/cm$^3$ (where 
$\rho_{\mathrm{drip}}\simeq$4.2$\times$10$^{11}$g/cm$^3$ 
is the neutron-drip density).
For the deeper inner-crust layers ($\rho > 10^{12}~{\rm g/cm^{3}}$), large 
neutron-to-seed ratios  drive the nuclear flow into the very heavy-mass region, 
leading to multiple fission recycling. As a consequence, the resulting abundance 
distribution becomes independent of the initial conditions, especially of the 
initial density. It was found to be in close agreement with the
solar distribution for $A>140$ nuclei \citep{go05,ar07}.

Different approaches were taken to nucleosynthesis calculations for merger ejecta
by \citet{frei99,metzger10} and \citet{roberts11}. In their calculations, while the 
density evolution of the mass elements was adopted from hydrodynamical simulations,
both the initial neutron enrichment and the temperature 
history were considered as free parameters. In particular, $Y_e$ was chosen in 
order to obtain, after decompression, an r-abundance distribution as close as 
possible to the solar distribution. This led to values of $Y_e=0.1$ 
\citep{frei99} or 0.2 \citep{roberts11}, corresponding to relatively near-surface
layers of the inner crust and to nuclear flows that are not subject to multiple 
fission cycles. In our simulations, most of the ejecta mass originates from the 
deep layers of the inner crust so that any contribution from near-surface layers 
remains minor. In addition, previous studies assumed that the nucleosynthesis 
is independent of the initial temperature of the ejected material, while we find 
that the initial temperature may not only affect the initial composition, but 
potentially also the nucleosynthesis (see Sect.~\ref{sect_in}) so that special 
attention should be paid to the detailed temperature history of the ejected 
material prior to its free expansion.

The work presented here is based on recent three-dimensional relativistic 
simulations of NS-NS mergers to determine the nucleosynthesis-relevant conditions 
of the ejected matter. The hydrodynamical model is described in Sect.~\ref{sect_hydro}.
Section \ref{sect_in} presents the nucleosynthesis results for two binaries, 
a symmetric NS-NS system and an asymmetric one. 
In contrast to previous studies, detailed information about the density, 
$Y_e$, and entropy evolution of the ejecta is extracted from 
the hydrodynamical simulations and included in the network calculations. The 
expected electromagnetic emission that is powered by radioactive decays following 
the heavy-element nucleosynthesis is estimated by a simple, approximative 
light-curve model in Sect.~\ref{sect_lc}. 
Conclusions are drawn in Sect.~\ref{sect_conc}.

%------------------------------------------------
\begin{figure*}
\begin{center}
\includegraphics[width=2.\columnwidth]{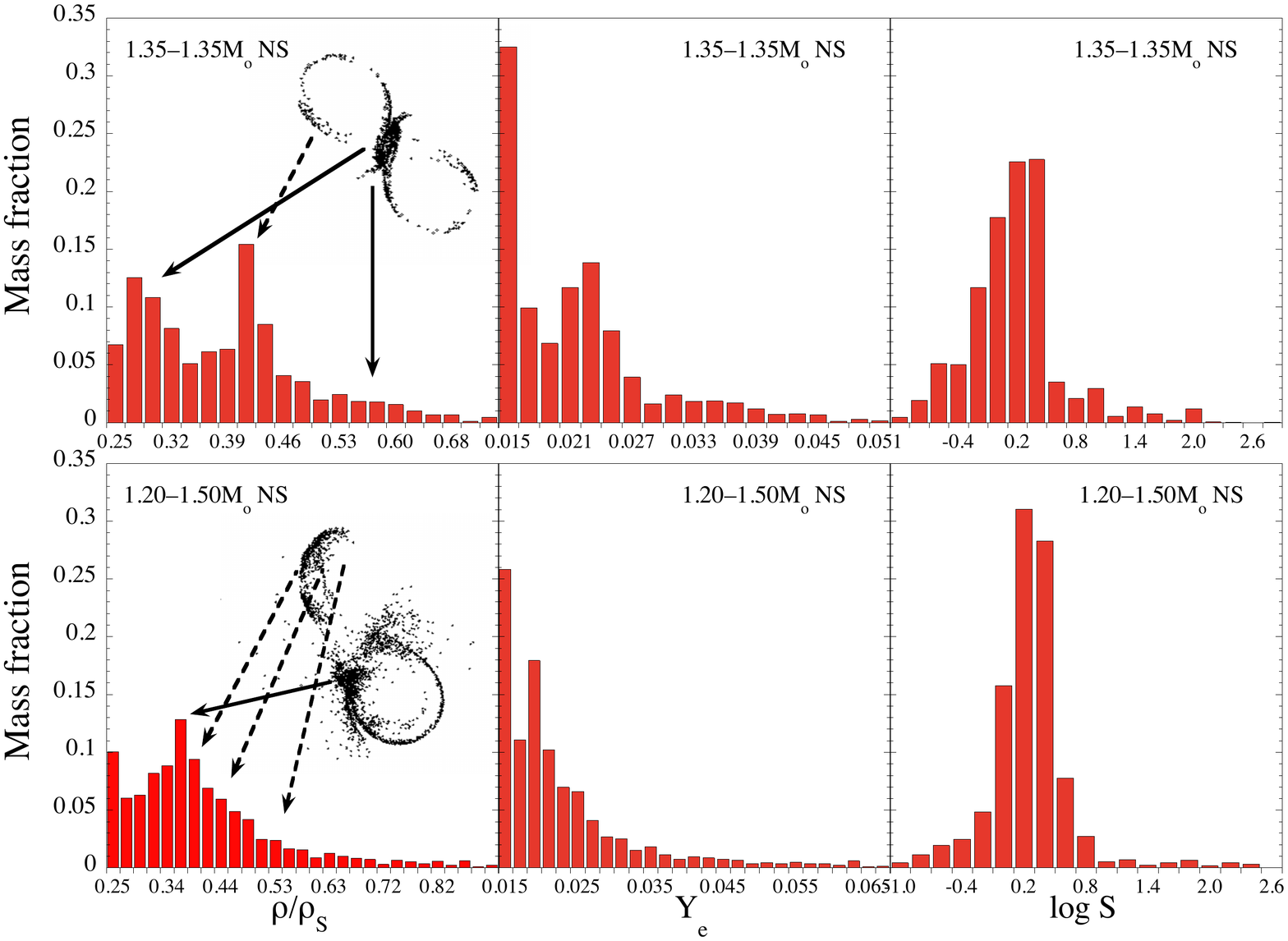}
\end{center}
\vskip -0.5cm
\caption{Histograms of fractional mass distribution of the ejecta
for the 1.35--1.35\Msun\ NS merger (upper row) and the
1.2--1.5\Msun\ binary (lower row) as functions of density $\rho$
(relative to the saturation density $\rho_S\simeq 2.6\times$10$^{14}{\rm g/cm^3}$;
left) and of electron fraction $Y_e$ (middle) that the ejected matter
had at its initial NS location prior to merging.
The right panels show the fractional mass distributions as functions of the
final entropy $S$ per nucleon when the matter starts its free expansion.
In the insert on the left panels the dots mark positions of
mass elements that get ejected later. The locations are given in
projection on the orbital plane at the time when the stellar collision
begins. }
\label{fig_histo}
\end{figure*}
%--------------------------------------------------

\section{HYDRODYNAMICAL MODELS}
\label{sect_hydro}

Our NS--NS merger simulations were performed with a general relativistic Smoothed 
Particle Hydrodynamics scheme \citep{oech07,bauswein10} representing the fluid by 
a set of particles with constant rest mass, whose hydrodynamical properties were 
evolved according to Lagrangian hydrodynamics, keeping $Y_e$ of fluid 
elements fixed. The Einstein field equations were
solved assuming a conformally flat spatial metric. Binaries with two mass ratios 
were modeled, namely symmetric 1.35\Msun--1.35\Msun\ and asymmetric 
1.2\Msun--1.5\Msun\ systems, both with a resolution of about 550,000 particles. 
The 1.35\Msun--1.35\Msun\ case is of particular interest since, according to 
population synthesis studies and pulsar observations, it represents the most 
abundant systems \citep{bel08}. 

For the results presented here we used the \citet{shen98} equation of state (EOS),
which includes thermal effects that become important when the NSs collide. The
corresponding NSs have radii of $\sim$15\,km \citep[see][]{bauswein10}. To assess 
the EOS influence on gross properties of the ejecta, we also performed simulations 
with the \cite{lattimer91} LS220 EOS with an incompressibility modulus of 220~MeV
and a NS radius of $\sim$12\,km. Note 
that both of our EOSs are in agreement with the observation of a 2\,$M_\odot$ NS 
\citep{demorest10}. Since an accurate calculation of the temperature evolution in
the hydrodynamic simulations is hampered by high initial degeneracy and limited
numerical resolution, we postprocessed the temperature of each ejected fluid element
by assuming an adiabatic flow in the absence of shocks and increasing the specific
entropy (consistent with the shock-jump conditions) when a shock was detected.

In the case of the symmetric NS binary, about 3$\times 10^{-3}$\Msun\ 
are found to become gravitationally unbound, whereas about 6$\times 10^{-3}$\Msun\ are 
ejected from the asymmetric system (using the LS220 EOS, we obtain an
ejecta mass of $\sim$2$\times 10^{-3}$\Msun\ and 6$\times 10^{-3}$\Msun\, respectively). 
The ejected ``particles'' (i.e., mass elements) originate mostly from two different
regions in the inner crust of the initial stars. For the symmetric model $\sim$75\% 
of the material are squeezed out from the contact interface of the NSs. The 
remaining 25\% are ejected from the near-surface regions close to the orbital plane,
see inset of Fig.~\ref{fig_histo}, where the arrows indicate by trend which initial 
densities the majority of fluid elements corresponds to. Note that the ``contact'' 
interface of the asymmetric merger is different from the symmetric case. While 
in the latter the two NSs collide in a shearing way and violently along a broad 
area, the impact of the tidally stretched lower-mass companion in the asymmetric 
case happens in a more grazing way such that a conical ``nose'' of the lower-mass 
star digs off matter from the higher-mass object. For this reason, in the asymmetric 
system the contact region preferentially contributes to ejecta with low initial
densities. As seen in Fig.~\ref{fig_histo}, all ejecta have low $Y_e$ ranging
between 0.015 and 0.050 (about 30\% of the mass have $Y_e \sim 0.015$). In the 
asymmetric case a small amount of ejected material starts out from densities above
2$\times 10^{14}{\rm g/cm^3}$ and has $Y_e$-values up to 0.07.
For the first 27\,ms the density history is consistently followed by the numerical 
simulation. Afterwards the escaping ejecta are assumed to expand freely with constant 
velocity. The radii of the ejecta clumps thus grow linearly with time $t$ and 
consequently their densities drop like $1/t^3$. Note that due to a lack
of resolution the dynamics and mass of unbound crust material with initial densities
$\rho \lesssim 10^{13}{\rm g/cm^3}$ cannot be reliably calculated in the 
hydrodynamic model. However, compared to the $10^{-3}$--$10^{-2}$\Msun\ of
inner crust matter its contribution to the ejecta remains small.

The ejected matter is initially cold, but most of it gets shock-heated during the
ejection to temperatures above 1\,MeV. Its composition is then determined by 
nuclear statistical equilibrium (NSE). When the drip density is reached during
expansion, most of the matter has cooled below 1~MeV and the NSE composition has 
frozen out. As soon as the temperature has dropped below $10^{10}$~K, further
changes of the composition are followed by a full network calculation (as detailed
below), and the temperature evolution is determined on the basis of the laws of 
thermodynamics, allowing for possible nuclear heating through $\beta$-decays, 
fission, and $\alpha$-decays, as described in \citet{meyer89}.

\section{NUCLEOSYNTHESIS}
\label{sect_in}

\begin{figure}
\begin{center}
\includegraphics[width=1.0\columnwidth]{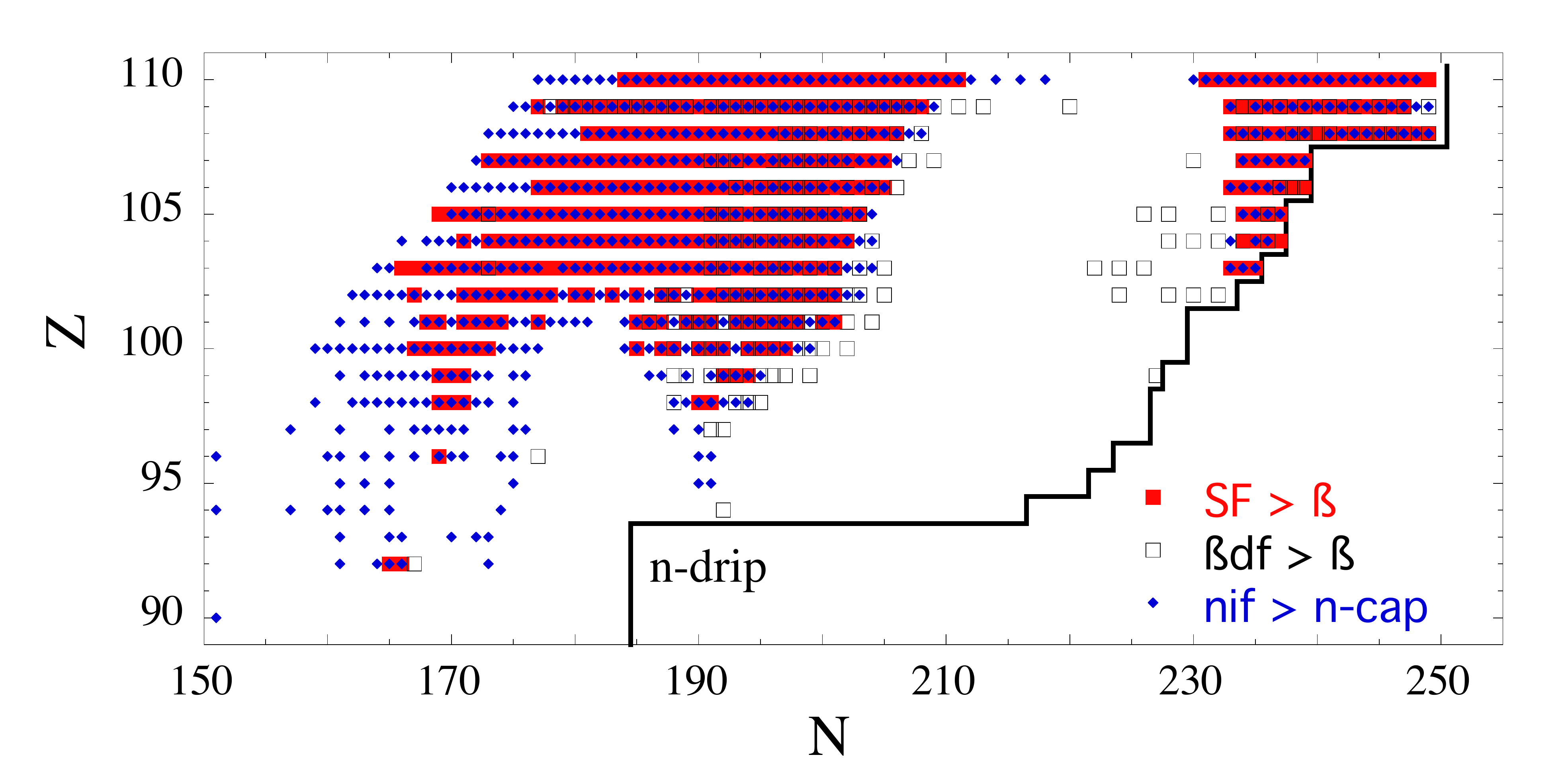}
\end{center}
\vskip -0.5cm
\caption{Representation of dominant fission regions in the $(N,Z)$ plane. 
Nuclei for which spontaneous fission is estimated to be faster than 
$\beta$-decays are shown by full squares, those for which $\beta$-delayed 
fission is faster than $\beta$-decays by open squares, and those for 
which neutron-induced fission is faster than radiative neutron 
capture at $T=10^9\,$K by diamonds.}
\label{fig_fission}
\end{figure}

The reaction network includes all 5000 species from protons up to Z=110 lying 
between the valley of $\beta$-stability and the neutron-drip line. All fusion 
reactions on light elements that play a role when the NSE freezes out are 
included in addition to radiative neutron captures and photodisintegrations. 
The reaction rates on light species are taken from the NETGEN library, which 
includes all the latest compilations of experimentally determined 
reaction rates \citep{nacre2}. Experimentally unknown reactions are estimated 
with the TALYS code \citep{go08} on the basis of the HFB-21 nuclear mass model 
\citep{gcp10}. On top of these reactions, fission and $\beta$-decays are also 
included, i.e neutron-induced fission, spontaneous fission, $\beta$-delayed 
fission, photofission, as well as $\beta$-delayed neutron emission. The  
$\beta$-decay processes are taken from the updated version of the 
Gross Theory \citep{tachibana90} based on the HFB-14 $Q$-values, whereas all 
fission processes are estimated on the basis of the HFB-14 fission 
path and the full calculation of the corresponding barrier penetration 
\citep{go09}. The main fission region is illustrated in Fig.~\ref{fig_fission}. 
The fission fragment distribution is taken from \cite{kodoma75}, and the 
fragment mass- and charge-asymmetry are derived from the HFB-14 prediction of 
the left-right asymmetry at the outer saddle point.
Due to the specific initial conditions of high neutron densities (typically 
$N_n\simeq 10^{33-35} {\rm cm^{-3}}$ at the drip density), the nuclear flow 
during most of the neutron irradiation will follow the neutron-drip line. 
For these nuclei at $T \gtrsim 2$--3$\times 10^9$~K,
(n,2n) and (2n,n) reactions are faster than
($\gamma$,n) and (n,$\gamma$) reactions and must be included 
in the reaction network. The (n,2n) rates are estimated with 
the TALYS code and the reverse rates from detailed balance 
expressions. 

%------------------------------------------------
%--------------------------------------------------

\begin{figure}
\begin{center}
\includegraphics[width=1.0\columnwidth]{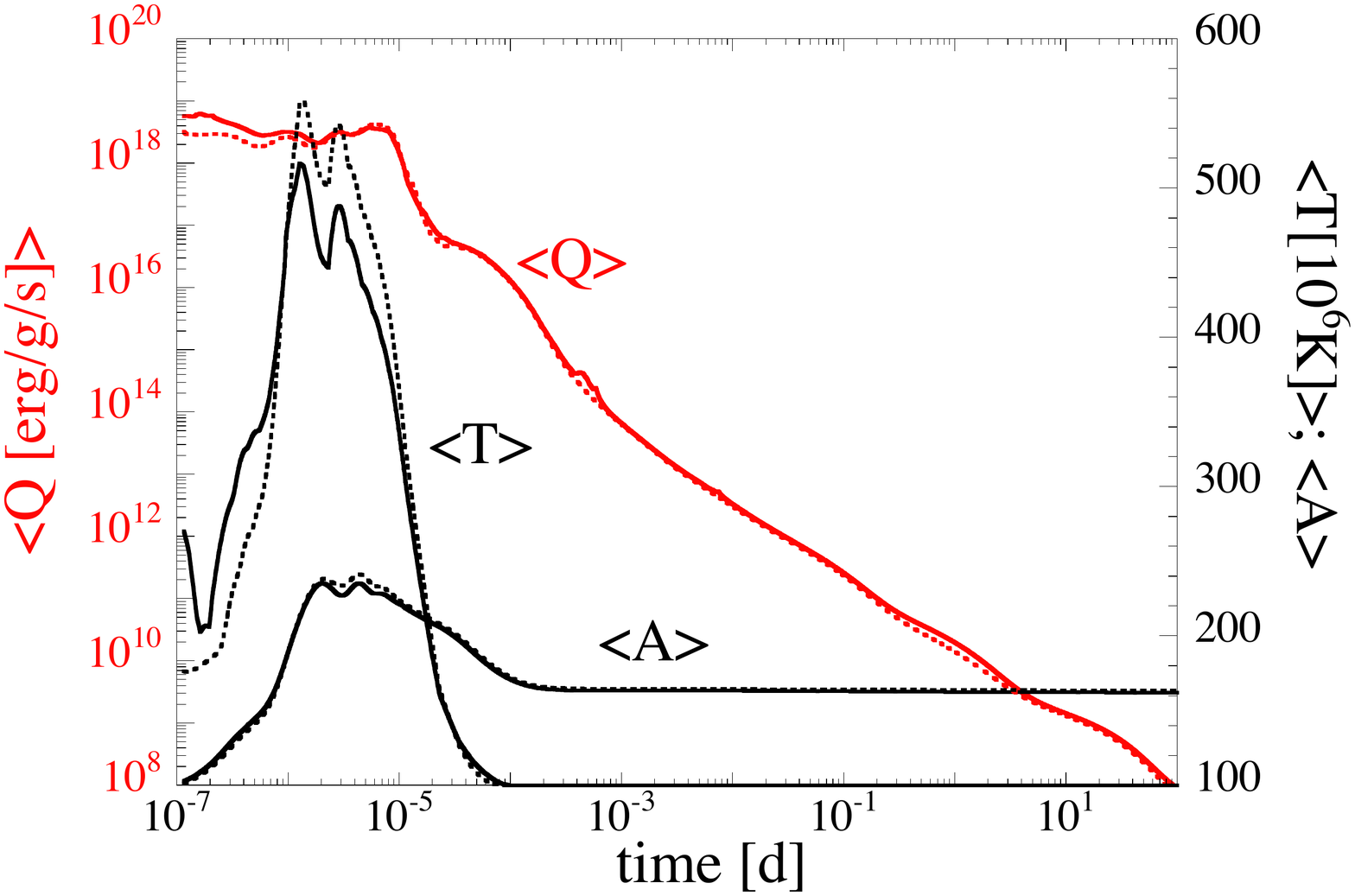}
\end{center}
\vskip -0.5cm
\caption{Time evolution of the total radioactive heating rate per
unit mass, $\langle Q\rangle$, mass number $\langle A\rangle$,
and temperature $\langle T\rangle$ (all
mass-averaged over the ejecta) for the 1.35--1.35\Msun\
(solid lines) and 1.2--1.5\Msun\ (dotted lines) NS mergers. }
\label{fig_QTA}
\end{figure}

For drip-line nuclei with $Z\ge 103$, fission becomes efficient 
(Fig.~\ref{fig_fission}) and recycling takes place two to three times 
before the neutrons are exhausted, as shown in Fig.~\ref{fig_QTA} 
by the time evolution of the mass number $\langle A\rangle$ mass-averaged 
over all the ejecta.
After several hundred ms, when neutrons get exhausted by captures 
($N_n \sim 10^{20}$~cm$^{-3}$), n-captures and $\beta$-decays compete 
on similar timescales and fashion the final abundance pattern before the
nuclear flow becomes dominated by $\beta$-decays 
(as well as fission and $\alpha$-decays for the heaviest species)
back to the stability line. The 
average temperature remains rather low during the late neutron irradiation, 
around 0.5\,GK (Fig.~\ref{fig_QTA}) so that photoreactions do not play a 
major role.  

\begin{figure}
\begin{center}
\includegraphics[width=1.0\columnwidth]{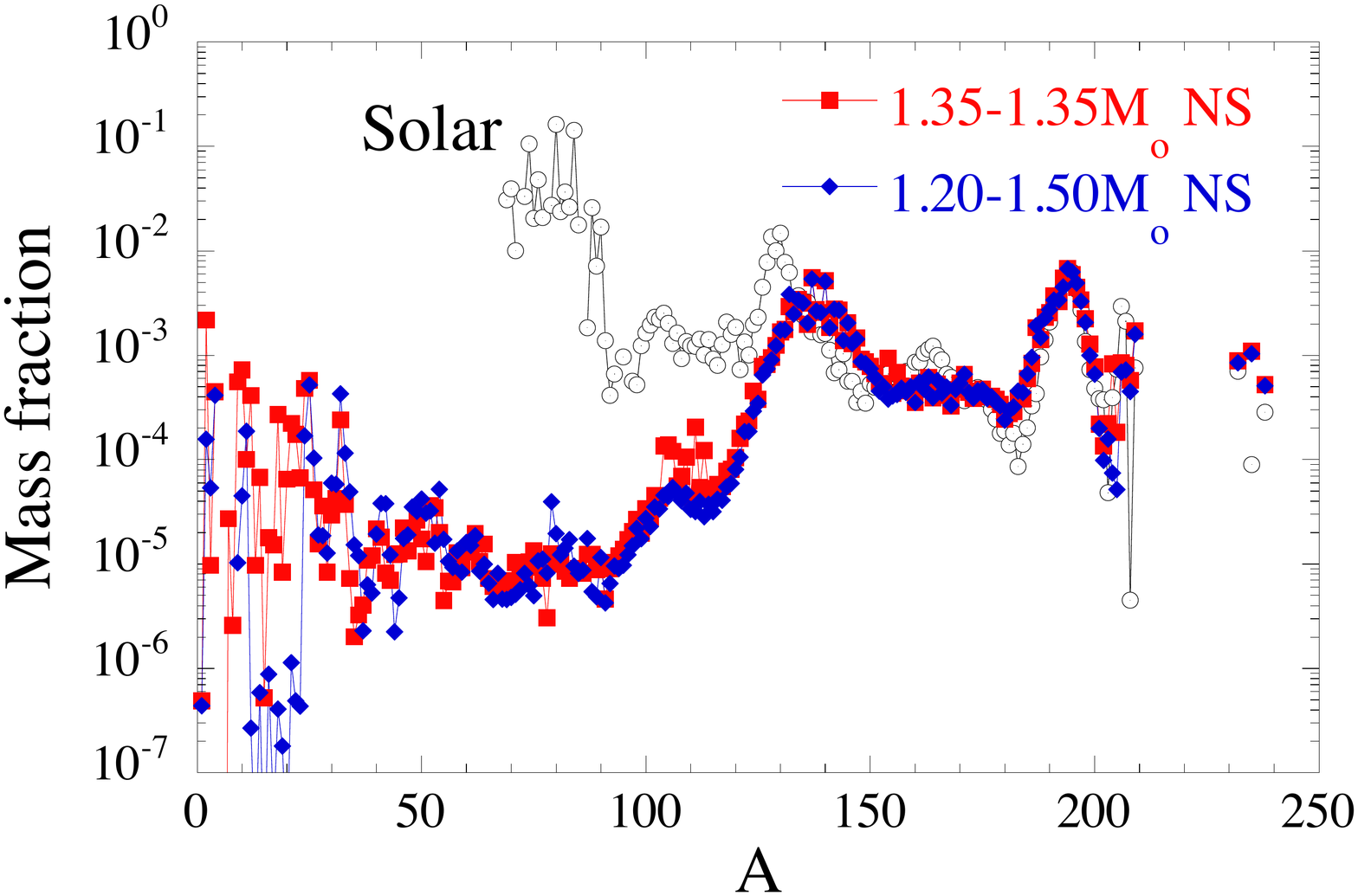}
\end{center}
\vskip -0.5cm
\caption{Final nuclear abundance distributions of the ejecta from 
1.35--1.35\Msun\ (squares) and 1.2--1.5\Msun\ (diamonds) NS mergers 
as functions of atomic mass. The distributions are normalized to the 
solar r-abundance distribution (dotted circles). }
\label{fig_ab}
\end{figure}

The final mass-integrated ejecta composition is shown in 
Fig.~\ref{fig_ab}. The $A=195$ abundance peak related to the $N=126$ shell 
closure is produced in solar distribution and found to be almost insensitive 
to all input parameters such as the initial abundances, the expansion 
timescales, and the adopted nuclear models. In contrast, the peak around 
$A=140$ originates exclusively from the fission recycling, which
takes place in the $A\simeq 280$--290 region at the time all neutrons 
have been captured. These nuclei are predicted to fission symmetrically
as visible in Fig.~\ref{fig_ab} by the $A\simeq 140$ peak corresponding to 
the mass-symmetric fragment distribution. It is emphasized that 
significant uncertainties still affect the prediction of fission 
probabilities and fragment distributions so that the exact strength 
and location of the $A\simeq 140$ fission peak (as well as the possible 
$A=165$ bump observed in the solar distribution) depend on the adopted 
nuclear model. 

While most of the matter trajectories are subject to a density and 
temperature history leading to the nuclear flow and abundance distribution
described above, some mass elements can be shock-heated at 
relatively low densities. Typically at
$\rho>10^{10}~{\rm g/cm^3}$ the Coulomb effects shift the NSE 
abundance distribution towards the high-mass region \citep{go11}, 
but at lower densities, the high temperatures lead to the 
photodissociation of all the medium-mass seed nuclei into neutrons 
and protons. Nucleon recombination may occur during the 
decompression provided the expansion timescale of the trajectories 
is long enough. For a non-negligible amount of ejected material, this 
recombination is indeed inefficient so that light species (including 
D and $^4$He) are also found in the ejecta (Fig.~\ref{fig_ab}). 
The final yields of $A<140$ nuclei remain, however, small and are not 
expected to contribute to any significant enrichment of the interstellar 
medium compared to the heavier r-elements.

%------------------------------------------------
%--------------------------------------------------

%------------------------------------------------
%--------------------------------------------------

\section{ELECTROMAGNETIC COUNTERPARTS}
\label{sect_lc}

Radioactive power through $\beta$-decays, fission processes as well as
late-time $\alpha$-decays will heat the expanding ejecta and make them
radiate as a ``macro-nova'' \citep{kulkarni05}
or ``kilo-nova'' \citep{metzger10} 
associated with the ejection of nucleosynthesis 
products from the merger \citep{li98}.
The time evolution of the corresponding total mass-averaged energy
release rate available for heating the ejecta (i.e., energy escaping
in neutrinos is not considered) is
plotted in Fig.~\ref{fig_QTA} for both the 1.35--1.35\Msun\ and 1.2--1.5\Msun\ 
binaries. While $\langle Q(t)\rangle$ and the average temperature evolution 
differ only slightly  
between both NS-NS systems, the ejecta masses $M_\mathrm{ej}$ 
and mass-averaged expansion velocities $v_\mathrm{exp}$ differ considerably.
While we find for the symmetric system 
$v_\mathrm{exp} \approx 0.31c$ ($c$ being the speed of light)
and $M_\mathrm{ej} \approx 3\times 10^{-3}\,M_\odot$, corresponding
to a total heating energy of $E_\mathrm{heat} \approx 2\times 10^{49}\,$erg 
or 3.4\,MeV/nucleon, the numbers for the 
asymmetric case are $v_\mathrm{exp} \approx 0.23c$, 
$M_\mathrm{ej} \approx 6\times 10^{-3}\,M_\odot$ and 
$E_\mathrm{heat} \approx 4\times 10^{49}\,$erg 
(again 3.4\,MeV/nucleon)\footnote{In the simulations with the LS220 EOS
we obtain $v_\mathrm{exp} \approx 0.28c$ for the symmetric and 
$v_\mathrm{exp} \approx 0.24c$ for the asymmetric binary.}.
This must be expected to lead to significant differences in the
brightness evolution of the kilo-nova because its peak
bolometric luminosity scales with 
$L_\mathrm{peak}\propto v_\mathrm{exp}^{1/2} M_\mathrm{ej}^{1/2}$ and,
for free expansion ($v_\mathrm{exp}$ = const), is reached on a time-scale
$t_\mathrm{peak} \propto v_\mathrm{exp}^{-1/2} M_\mathrm{ej}^{1/2}$
\citep{metzger10,arnett82}.

We calculate an approximation of the light-curves of such events by
employing a semi-analytic, simplified one-zone model of 
\citet{kulkarni05} \citep[see also][]{arnett82,li98}, assuming
that asymmetries of the emission remain modest
\citep[cf.][]{roberts11}. The EOS includes photons as well
as nuclei and electrons, whose recombination fraction with decreasing
temperature is approximated by that of $^{56}$Ni \citep{arnaud85}.
The opacity $\kappa = 0.4\,Z/A$\,cm$^2$g$^{-1}$ for Thomson 
scattering is used, taking
mass-averaged $Z$ and $A$ from the nucleosynthesis yields
(Fig.~\ref{fig_QTA}) and ignoring electron recombination 
as suggested by \citet{roberts11}.

\begin{figure}
\begin{center}
\includegraphics[width=1.0\columnwidth]{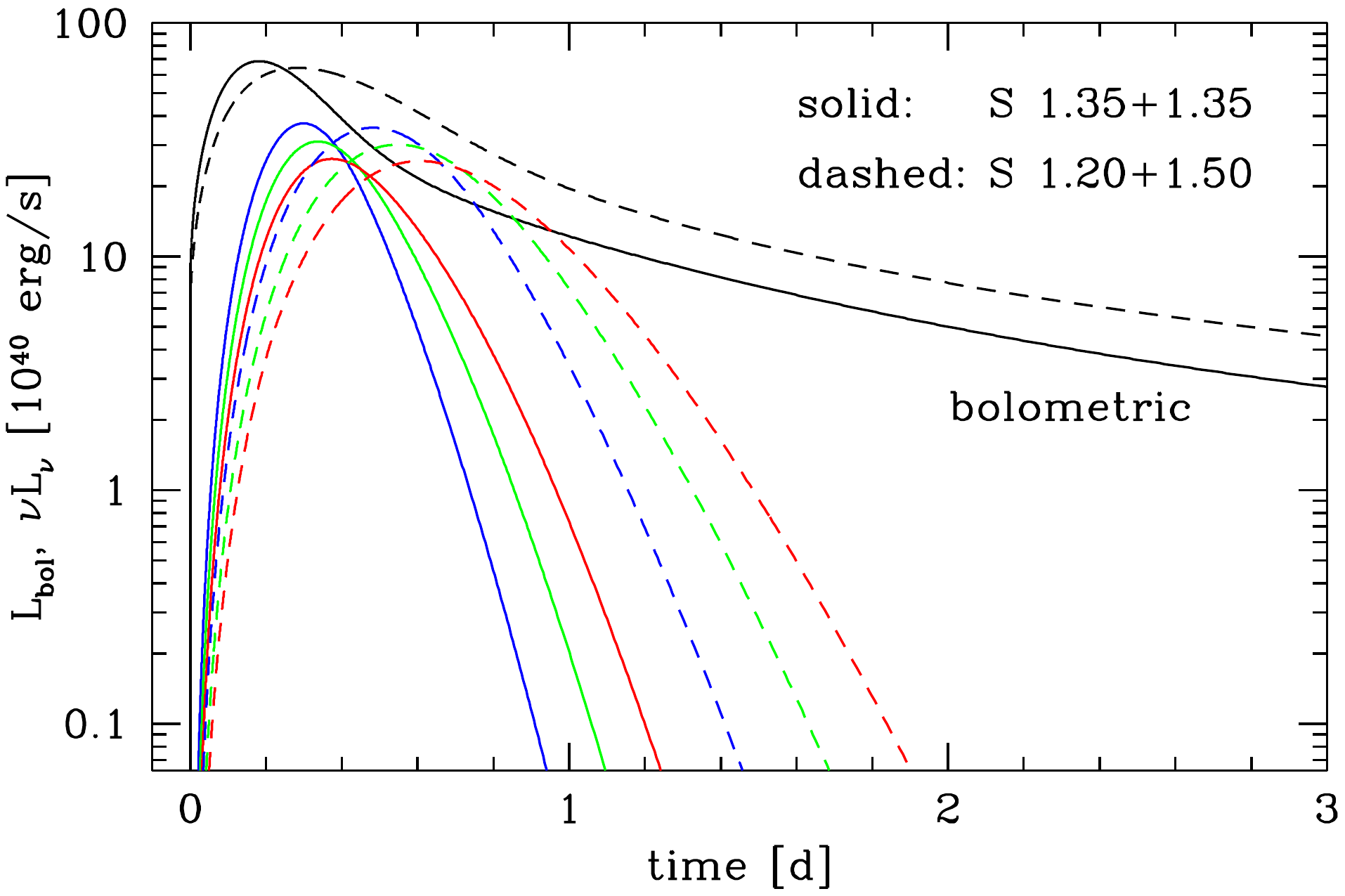}
\end{center}
\vskip -0.5cm
\caption{Photon luminosities of the expanding NS merger ejecta
caused by radioactive decay heating
for the 1.35--1.35\Msun\ (solid lines) and 1.2--1.5\Msun\ 
(dashed lines) binaries. The upper, long-duration lines are the bolometric
luminosities, the sequences of short-duration peaks correspond to the emission
in the blue, visual, and red wavebands (at wavelengths of 445, 551, 658\,nm; 
from left to right).
}
\label{fig_lum}
\end{figure}

Our results for the bolometric light curves $L_\mathrm{bol}(t)$ 
and the B-, V-, and R-band luminosities $\nu L_{\nu}(t)$
are displayed in Fig.~\ref{fig_lum}. While we expect 
significant emission in the chosen wavelength bands up to 
$(\nu L_\nu)_\mathrm{peak}\approx 4\times 10^{41}\,$erg\,s$^{-1}$ 
for about one day
in the case of a 1.35--1.35\Msun\ merger, the 1.2--1.5\Msun\ system
produces sizable BVU-radiation nearly twice as long.

%------------------------------------------------
%--------------------------------------------------

\section{CONCLUSIONS}
\label{sect_conc}

Using relativistic NS merger models to determine the nucleosynthesis-relevant
conditions self-consistently, we confirm that decompressed NS matter ejected
dynamically during the stellar collision and shortly afterwards is an extremely
promising site for robust, strong r-processing. Matter from the inner crust of
the coalescing NSs, which dominates the ejecta by far, produces an r-abundance 
distribution very similar to the solar one for nuclei with $A>140$.
Nuclei with $A<140$ with solar distribution could originate
from the outer crust \citep{go11}, but too little of such matter gets ejected to
explain the solar proportion of light-to-heavy r-process material. However,
significant amounts of $A < 140$ nuclei might be produced in the outflow of a
BH-torus system formed after the NS merger (Wanajo \& Janka 2011).
 
The underlying nuclear mechanisms differ significantly from those at action in 
SN scenarios. In particular fission plays a major role in recycling 
heavy material. The similarity between predicted and solar abundance patterns 
as well as the robustness of the prediction against variations of input parameters,
which we have shown in \cite{go05}, have demonstrated here in comparison of symmetric
and asymmetric NS-NS mergers, and will further elaborate on in a forthcoming paper,
make this site one of the most promising, deserving further exploration concerning
hydrodynamics, galactic chemical evolution, nucleosynthesis, nuclear physics,
and astronomical consequences.

Fully relativistic simulations including neutrino transport and magnetic fields,
with good resolution of the inner (and outer) crust layers of the merging NSs
assured by adaptive refinement, are needed to corroborate the ejecta conditions 
found in our work. With our yield of $\sim$(3--5)$\times 10^{-5}\,M_\odot$ per 
event of $^{151,153}$Eu, a nearly-pure r-process element, 
the origin of all galactic Eu from NS-NS mergers would require an event rate of 
$\sim$(2--3)$\times 10^{-5}$yr$^{-1}$, fully compatible with present best
estimates \citep[see, e.g.,][]{bel08}. However, also mass loss during the evolution
of the merger remnant from a hypermassive NS to a massively accreting
BH-torus system as well as NS-BH mergers with higher ejecta masses might add to
the production (e.g., Wanajo \& Janka 2011, Caballero et al.\ 2011) and deserve
more detailed investigations.
From the nuclear point of view, this site also implies new challenges because it
involves the formation of neutron-drip nuclei, for which $\beta$-decay, neutron 
capture, and fission rates need to be determined. Astronomically, the discovery of
electromagnetic radiation from kilo-nova events \citep{roberts11,metzger10}
could mean the first in-situ observation of freshly produced r-process material.

\acknowledgments

H.-T.J.\ thanks M.\ Kromer and S.\ Hachinger for discussions.
DFG grants SFB/TR7, SFB/TR27, and EXC~153,
computing at LRZ and RZG, and support of F.N.R.S.\ and
``Actions de recherche concert\'ees (ARC)'' from the 
``Communaut\'e fran\c caise de Belgique'' are acknowledged.

%\clearpage


\begin{thebibliography}{}

\bibitem[Argast et al. (2004)]{argast04} Argast, D.,   Samland, M.,  Thielemann, F.-K. \& Qian, Y. 2004, \aap, 416 997
\bibitem[Arnaud \& Rothenflug (1985)]{arnaud85} Arnaud, M. \& Rothenflug, R. 1985, \aaps, 60, 425
\bibitem[Arnett (1982)]{arnett82} Arnett, W.D. 1982, \apj, 253, 785
\bibitem[Arnould et al. (2007)]{ar07} Arnould, M., Goriely, S., \& Takahashi, K. 2007,  \physrep, 450, 97
\bibitem[Bauswein et al. (2010)]{bauswein10}  Bauswein, A.,  Janka, H.-T., \& Oechslin, R. 2010, \prd, 82,  084043
\bibitem[Belczynski et al. (2008)]{bel08}{Belczynski}, K., {O'Shaughnessy}, R.,  {Kalogera}, V., {Rasio}, F., {Taam}, R.~E. \& {Bulik}, T. 2008, \apjl, 680, L129
\bibitem[Caballero et al. (2011)]{caball11} Caballero, O.L., McLaughlin, G.C., \& Surman, R. 2011, arXiv:1105.6371; \apj, submitted 
\bibitem[Demorest et al. (2010)]{demorest10}{Demorest}, P.~B., {Pennucci}, T.,  {Ransom}, S.~M., 
	{Roberts}, M.~S.~E., \&  {Hessels}, J.~W.~T. 2010 \nat, 467, 1081
\bibitem[Fischer et al. (2010)]{fisch10} Fischer, T., Whitehouse,  S. C., Mezzacappa, A., Thielemann,  F.-K., \& Liebend\"orfer,  M.  2010, \aap, 517, A80
\bibitem[Freiburghaus et al. (1999)]{frei99}  Freiburghaus, C.,  Rosswog, S., Thielemann, F.-K. 1999, \apj, 525,  L121
\bibitem[Goriely et al. (2005)]{go05} Goriely, S., Demetriou, P.,  Janka, H.-T. \& Pearson, J.M. 2005, \nphysa, 758,  587c
\bibitem[Goriely et al. (2008)]{go08} Goriely, S., Hilaire,  S., \& Koning, A.J. 2008,  \aap, 487,  76
\bibitem[Goriely et al. (2009)]{go09} Goriely, S., Hilaire, S., Koning, A.J., Sin, M. \& Capote, R. 2009, \prc, 79,  024612
\bibitem[Goriely et al. (2010)]{gcp10} Goriely, S., Chamel, N., \& Pearson J.M., 2010,  \prc, 82,  035804
\bibitem[Goriely et al. (2011)]{go11} Goriely, S., Chamel, N., Janka H.-T., \& Pearson, J.M. 2011, \aap, 531, A78
%  \bibitem[Granot et al. (2006)]{granot06} Granot, J., Ramirez-Ruiz, E., Taylor, G.B., et al. 2006 \apj, 638, 391
\bibitem[Hoffman et al. (2008)]{hoffman08} Hoffman, R.D., M\"uller, B., \& Janka, H.-T., 1999,  \apjl, 676, L127
\bibitem[H\"udepohl  et al. (2010)]{hudepohl10} H\"udepohl, L., M\"uller, B., Janka H.-T., Marek., A., \& Raffelt, G.G. 2010, \prl, 104, 251101
\bibitem[Janka et al. (1999)]{janka99} Janka, H.-T., Eberl, T., Ruffert, M., \& Fryer, C.L. 1999,  \apjl, 527,  L39 
\bibitem[Janka et al. (2008)]{janka08} Janka, H.-T., M\"uller, B., Kitaura, F.S., \& Buras, R. 2008, \aap, 485,  199
\bibitem[Kodoma \& Takahashi (1975)]{kodoma75} Kodoma, T. \& Takahashi, K. 1975,  \nphysa, 239,  489
\bibitem[Kulkarni (2005)]{kulkarni05} Kulkarni, S.R. 2005, arXiv:astro-ph/0510256
\bibitem[Lattimer et al. (1977)]{lattimer77} Lattimer, J.M., Mackie, F., Ravenhall, D.G., \& Schramm, D.N. 1977,  \apj, 213,  225
\bibitem[Lattimer et al. (1991)]{lattimer91} Lattimer, J.M. \& Swesty, F.~D. 1991, \nphysa, 535, 331
\bibitem[Li \& Paczy{\'n}ski (1998)]{li98} Li, L.-X. \& Paczy{\'n}ski, B. 1998, \apjl, 507, L59
\bibitem[Metzger et al. (2010)]{metzger10} Metzger, B.D., Martinez-Pinedo, G., Darbha, S., et al. 2010, \mnras, 406, 2650
\bibitem[Meyer (1989)]{meyer89} Meyer, B.S. 1989, \apj, 343, 254
\bibitem[Oechslin et al. (2007)]{oech07} Oechslin, R.,  Janka, H.-T., \& Marek,  A. 2007,  \aap, 467,  395
\bibitem[Roberts et al. (2010)]{roberts10} Roberts, L.F., Woosley, S.E., \& Hoffman, R.D. 2010, \apj, 722, 954 
\bibitem[Roberts et al. (2011)]{roberts11} Roberts, L.F., Kasen, D., Lee, W.H., \& Ramirez-Ruiz, E. 2011, \apj, in press
\bibitem[Rosswog et al. (2004)]{ros04} Rosswog, S., Speith, R., \& Wynn, G.A. 2004,  \mnras, 351,  1121
\bibitem[Shen et al. (1998)]{shen98}  Shen, H., Toki, H., Oyamatsu, K., \& Sumiyoshi, K. 1998,  \nphysa, 637,  435
\bibitem[Sneden et al. (2008)]{sneden08} Sneden, C., Cowan, J.J., \& Gallino, R. 2008, ARA\&A, 46, 241
%  \bibitem[Sumiyoshi et al. (1998)]{sumi98} Sumiyoshi, K., Yamada, S., Suzuki, H., \& Hillebrandt, W. 1998, \aap, 334, 159
\bibitem[Tachibana et al. (1990)]{tachibana90} Tachibana, T., Yamada, M., \& Yoshida, Y., 1990, Prog. Theor. Phys., 84,  641
\bibitem[Tsuruta \& Cameron (1965)]{tsuruta65} Tsuruta, S. \&  Cameron, A.G.W. 1965, Can. J. Phys., 43,  2056
\bibitem[Wanajo \& Janka (2011)]{wanajan11} Wanajo, S. \& Janka, H.-Th. 2011, arXiv:1106.6142; \apj, submitted
\bibitem[Wanajo et al. (2011)]{wanajo11} Wanajo, S., Janka, H.-Th., \& M\"uller, B., 2011,  \apjl, 726, L15
\bibitem[Xu et al. (2011)]{nacre2} Xu Y., Takahashi, K., Goriely, S., \& Arnould, M. 2011, in Frontiers in Nuclear Structure, Astrophysics and Reactions, eds. P. Demetriou et al., (AIP Conference) in press (see also {\it http://www.astro.ulb.ac.be/Netgen})

\end{thebibliography}
\end{document}